# Broadband microwave waveform generation with programmable chirp shapes via recirculating phase-modulated optical fiber loop controlled by low-speed electronics

Weiqiang Lyu[1,2], Huan Tian[1,2], Zhenwei Fu[1,2], Lingjie Zhang[1,2], Zhen Zeng[1,2], Yaowen Zhang[1,2], Zhiyao Zhang[1,2,*] & Yong Liu[1,2,*]

[1] State Key Laboratory of Electronic Thin Films and Integrated Devices, School of Optoelectronic Science and Engineering, University of Electronic Science and Technology of China, Chengdu 610054, P. R. China
[2] Research Center for Microwave Photonics (RC-MWP), School of Optoelectronic Science and Engineering, University of Electronic Science and Technology of China, Chengdu 610054, P. R. China
*Corresponding author: Zhiyao Zhang (email: zhangzhiyao@uestc.edu.cn); Yong Liu (email: yongliu@uestc.edu.cn)

Broadband microwave waveforms with programmable chirp shapes are captivating in numerous practical applications. Compared with electronic technology, photonic-assisted solutions exhibit excellent performance in bandwidth and flexibility, but still suffer from complex architecture and requirement of high-speed electronics. Besides, rapid manipulation of chirp shape is still a challenge in the scientific community. In this paper, we propose and demonstrate a novel concept for generating broadband microwave waveforms with programmable chirp shapes. This concept is realized on a simple fiber-optic platform involving a continuous-wave laser source, a recirculating phase-modulated optical fiber loop, and low-speed electronics with a sampling rate at the level of MS/s. Based on this method, chirped microwave waveforms with a bandwidth up to tens of GHz can be generated, where the chirp shape is identical to the low-frequency driving waveform of the recirculating phase-modulated optical fiber loop. In addition, all the parameters of the generated chirped microwave waveforms can be easily reconfigured in real time, including the bandwidth, the central frequency, and the temporal duration. In the experiment, broadband microwave waveforms with customized chirp shapes are generated, where the center frequency and bandwidth tuning ranges exceed 21 GHz, the temporal duration is tuned in the range of 9 ns to 180 ns, and the coherent time of the generated microwave waveform is larger than 100 μs. This simple fiber-optic platform paves a way to generate broadband microwave waveforms with user-definable chirp shapes, which can find applications in broadband radar systems, electronic warfare and wireless communications.

Broadband chirped microwave waveforms simultaneously providing features of large bandwidth and long duration are widely present in military and civilian applications, such as modern radar systems[1, 2], monitoring of human vital signs[3-5], wireless communication[6, 7], and optical vector analysis[8]. Chirped microwave signal sources with a large time-bandwidth product (TBWP) usually act as high-average-power probe signals, and the signal-to-noise ratio (SNR) can be significantly enhanced by performing pulse compression or matched filtering at receivers. The chirped microwave waveform generators applied in practical applications are required to offer excellent performance on size, weight, and power (SWaP) as well as rapidly reconfigure parameters (central frequency, bandwidth, time duration, chirp shape) of output chirped signals. Generally, broadband chirped microwave waveforms with flexible reconfigurability are synthesized in the electrical domain with the help of fast and high-frequency electronics, e.g., electronic arbitrary waveform generators (AWGs). Due to the constraint on the sampling rate of digital-to-analog converters, the attainable central frequency and bandwidth are limited to a few tens of GHz. Moreover, these high-performance devices are bulky, expensive, and with high power consumption, which hinders their widespread popularization in real-world applications.

Photonic solutions for chirped microwave waveform generation have been extensively investigated in the last 20 years thanks to the offered access to available bandwidth up to THz range, low-cost optical components, anti-electromagnetic interference, and light-weight photonic systems[9-11]. In addition, with the rapid advancement of photonic integration technology, a large number of mature photonic systems can be fabricated on chips, facilitating stabilization and miniaturization of photonic devices[12-14]. Up to now, numerous photonic approaches have been proposed to generate high-frequency and wide-bandwidth chirped microwave waveforms, e.g., direct space-to-time mapping[15-17], spectral shaping followed by frequency-to-time mapping[18-24], temporal pulse shaping[25-27]. Although many efforts have been devoted to developing their performance including the central frequency, the bandwidth, and the reconfigurability, these techniques still suffer from low SNR, small temporal duration, and limited TBWP (< 100). Another technique based on spectral line-by-line shaping is a finely spectral manipulation method for waveform synthesis[28-32], enabling the improvement of the SNR and the extension of the duration, but also faces a great challenge in TBWP (~ 100) limited by the number of the manipulable comb line. Some novel schemes based on a frequency-shifting loop can greatly increase the TBWP to above 1000[33-37], while only the linearly frequency-modulated or the frequency-stepped microwave waveforms can be generated. To improve the output bandwidth, a feasible approach is implemented by down-converting chirped optical waveforms to microwave domain via mixing them with a narrow-linewidth single-frequency laser[38, 39]. The performance of the output chirped microwave waveforms mainly depends on the chirped optical signals. Alternatively, the optical modulation technique can be utilized to boost the central frequency and the bandwidth of the external chirped microwave signals[40-44]. However, an electronic AWG is required to provide initial chirped microwave waveforms, leading to large complexity, big size, and high cost.

In the past few years, photonic generation of chirped microwave waveforms exhibiting a remarkable performance on TBWP is based on an optically injected semiconductor laser operated in period-one oscillation state[45-47]. Nevertheless, several stringent constraints on injection conditions should be satisfied to excite period-one oscillation[48], including polarization of injection light, frequency detuning, and injection power. Another inventive approach of Fourier domain mode locking (FDML) technique based on optoelectronic oscillator (OEO)[49-53] or laser[54] was proposed and demonstrated to be capable of generating chirped microwave waveforms with a large TBWP exceeding $10^5$ since it breaks the limitation of mode building time. The key component incorporated in OEO or laser is a high-quality-factor and fast-tuning filter with a tuning period or its multiple equal to the loop round-trip time. In this way, a fast frequency-scanning microwave or optical signal, i.e., frequency-chirped waveform, will be produced after stable oscillation. But for the generation of customized chirped microwave waveforms, a corresponding frequency-sweeping filter must be involved in this configuration, which increases the complexity of the photonic system to some extent. Besides, impeded by the existing techniques for building a frequency-sweeping filter[49, 50], the actual chirp shape of the microwave waveform generated from the FDML OEO/laser may be distorted compared with its desired form. Otherwise, a chirped microwave signal with an expectant chirp shape is inevitably applied to construct an accurately frequency-scanning track of the filter[53], resulting in an augmentation in cost and SWaP.

In this paper, we propose and demonstrate a novel concept for generating broadband microwave waveforms with customized chirp shapes using an extremely simple recirculating phase modulation configuration, avoiding the requirement of frequency-scanning filters/lasers, large dispersion mediums, or fast electronics. The proposed concept provides the excellently reconfigurable ability in the central frequency, the time duration, the bandwidth, and the time-frequency distribution of the generated chirped microwave waveforms as well as the strong mutual coherence. The most attractive characteristic of the proposed scheme is that the chirp shapes of the generated microwave waveforms, i.e., time-frequency distribution, are identical to the scaled forms of the MHz-range phase-modulated waveforms, which offers an extremely simple way to manipulate the chirp shape of a high-frequency and broadband microwave waveform by simply modifying the temporal waveform of a low-frequency signal. We have experimentally demonstrated the proposed concept in a fiber-based platform mainly constituted by a recirculating phase-modulated optical fiber loop. In our configuration, the central frequency and the bandwidth of the generated chirped microwave signals can be varied in a frequency range of about 22 GHz, restricted by the limited bandwidth of the optical detector, where the potential tuning range can reach several hundred GHz. The temporal duration is identical to the phase-modulated period. A temporal duration tuned from 9 ns to 180 ns by varying the modulation period is demonstrated. Besides, the manipulation of the time-frequency distribution is also demonstrated by generating linear/nonlinear chirped microwave signals, a frequency-hopping microwave signal, and two chirped microwave waveforms with customized chirp shapes. Finally, the calculated cross-correlation trace exhibits its coherent time exceeding 100 us. The robust reconfigurability offered by the proposed concept in all the parameters of the generated chirped microwave waveforms makes it more attractive in numerous practical applications, including ranging, spectroscopy, and imaging.

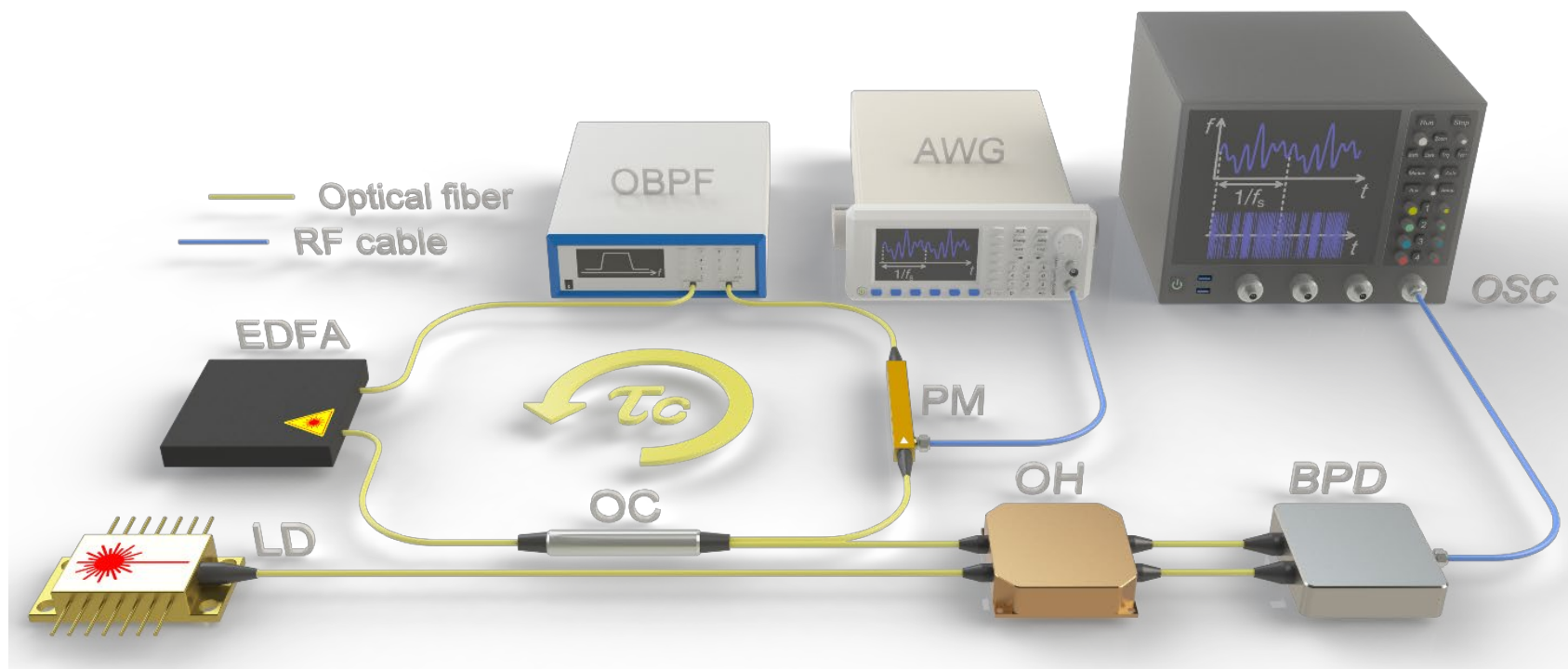

**Fig. 1**. Schematic of the proposed photonic architecture for generating microwave waveforms with customized chirp shapes. An optical loop (round-trip time: $\tau_c$) comprising an electro-optic phase modulator (PM), an optical bandpass filter (OBPF), an erbium-doped fiber amplifier (EDFA), and an optical coupler (OC) is utilized as an optical oscillator. A chirped optical waveform with a chirp shape identical to the phase-modulated waveform generated by an arbitrary waveform generator (AWG) will be produced from the PML. The output chirped optical signal is then down-converted to microwave domain by using an optical coherent detection device containing a 90° optical hybrid (OH) and a balanced photodetector (BPD), where a narrow-linewidth single-frequency laser from a laser diode (LD) acts as a local oscillator signal. The generated chirped microwave waveform is exhibited in an oscilloscope (OSC).

## Results

**Principle**. The reported photonic architecture for generating microwave waveforms with customized chirp shapes is based on a recirculating phase-modulated loop (PML) only seeded by an amplified spontaneous emission (ASE) noise at each roundtrip, as depicted in Fig. 1. The PML is composed of an electro-optic phase modulator (PM), an optical bandpass filter (OBPF) to control the frequency span of the recirculating optical filed, an erbium-doped fiber amplifier (EDFA) to supply the ASE noise and compensate for the loop losses, and an optical coupler (OC) to extract a fraction of the optical field. The PML is mainly described by two independent parameters: $\tau_c$, the round-trip time of the loop (or $f_c = 1/\tau_c$, its free spectral range (FSR)), and $f_s$, the repetition rate of the phase-modulated signal (or $\tau_s = 1/f_s$, its period).

In our configuration, the repetition rate of the phase-modulated signal is slightly detuned off an integer multiple of the loop FSR. We define $\Delta f = f_s - kf_c << f_c$ as the frequency detuning, where $k$ is a positive integer. Notice that $\Delta f$ can be positive or negative. In this way, the optical field in the PML is phase-modulated by a time-delayed signal $s(t\text{-}n\tau_c)$ in the roundtrip $n$. A tiny time parameter can be denoted as: $\tau_d = \tau_c - k\tau_s << \tau_c$. Since the phase-modulated signal is with a period of $\tau_s$, the following equation is satisfied: $s(t\text{-}n\tau_c) = s(t\text{-}n\tau_d)$. Here, suppose that the driving signal $s(t)$ can be considered as a constant within a minuscule time span of $\tau_d$. Thus, the recirculating phase modulation term experiencing $n$ roundtrips, $\exp\left(im\sum_{l=1}^{l=n}s(t\text{-}l\tau_d)\right)$, can be approximatively represented as: $\exp\left(\frac{im}{\tau_d}\int_{t-n\tau_d}^{t}s(u)du\right)$, where $m$ is the modulation index of the electro-optic PM. Besides, to ensure the self-oscillation of the PML seeded only by ASE noise $x_n(t)$, the net gain per roundtrip $\gamma$ should be slightly larger than 1. Then, the output optical field of the PML is the superposition of all the recirculating phase-modulated ASE noise, $E_0\sum_{n=1}^{n=N}\gamma^n x_n(t\text{-}n\tau_c)\exp\left(\frac{im}{\tau_d}\int_{t-n\tau_d}^{t}s(u)du\right)$, where $E_0$ and $N$ represent the amplitude of the ASE noise seeded per roundtrip and the total number of the roundtrips, respectively. As detailed in the Methods, the output of PML can be mathematically derived as the product of a chirped term, $\exp\left(\frac{imf_c f_s}{\Delta f}\int_0^t s(u)du\right)$, and a single-frequency signal with a frequency of $q_0f_c$, where $q_0$ is a positive integer. If the frequency detuning is set to be negative, the output optical field of the PML is a chirped waveform with a chirp shape identical to a scaled form of the phase-modulated waveform $s(t)$, a bandwidth of $|mf_cf_s / (\pi\Delta f)|$, and a central frequency of $q_0f_c$. In the other case of $\Delta f > 0$, the output chirp shape is proportionally identical to a reversed copy of $s(t)$, namely $-s(t)$. The central frequency $q_0f_c$ is determined by the mode competition between numerous intracavity modes, which can be tuned by adjusting the shape or central frequency of the OBPF. Finally, by virtue of an optical coherent receiver including a 90° optical hybrid and a balanced photodetector (BPD), the generated optical chirped signal is beat with a narrow-linewidth single-frequency laser at $f_{lo}$ to down-convert it to microwave domain, where the influence of the intra-comb beating is eliminated. As shown in the output chirped term, the phases of the output chirped microwave waveforms in different periods are entirely identical to each other, which ensures the mutual coherence of the output waveforms. Actually, several nonideal factors will degrade the coherence of the generated chirped microwave waveforms, including the phase noise of the phase-modulated signals, the stability of the loop length and the optical gain, the linewidth of the single-frequency laser used for frequency down-conversion.

The chirp shape of the generated chirped microwave waveforms can be customized by simply changing the temporal waveform of the low-speed phase-modulated signal, which gives a small SWaP access to obtain chirped microwave waveforms with a fast-reconfigurable chirp shape and a tunable central frequency. Besides, for a certain loop length, its bandwidth, $|mf_cf_s / (\pi\Delta f)|$, is dependent on the modulation index, the repetition rate of the phase-modulated signal, and the frequency detuning. The large modulation index can be achieved by use of the cascaded phase modulators. In addition, without the variation of the frequency detuning, the repetition rate $f_s$ can be manyfold enhanced to increase the bandwidth of the output chirped microwave waveforms. Nevertheless, the period of the generated chirped signals, $\tau_s = 1/f_s$, will be proportionally reduced, leading to a TBWP of $|mf_c / (\pi\Delta f)|$ in this architecture. Generally, the frequency detuning is a primary parameter used for controlling the bandwidth of the generated signals in the case of a fixed output period. Accordingly, the bandwidth of the OBPF should be adjusted to match the output bandwidth. In most cases based on fiber-optic platform, the ratio of the loop FSR to the frequency detuning, $f_c/\Delta f$, can reach several hundred or even one thousand, depending on the quality factor of the optical loop. Besides, the commercial PMs can easily meet the injection microwave power requirement for a modulation index of $2\pi$. As a result, the TBWP of chirped microwave waveforms produced from the fiber-based PML is about several hundred to a few thousand.

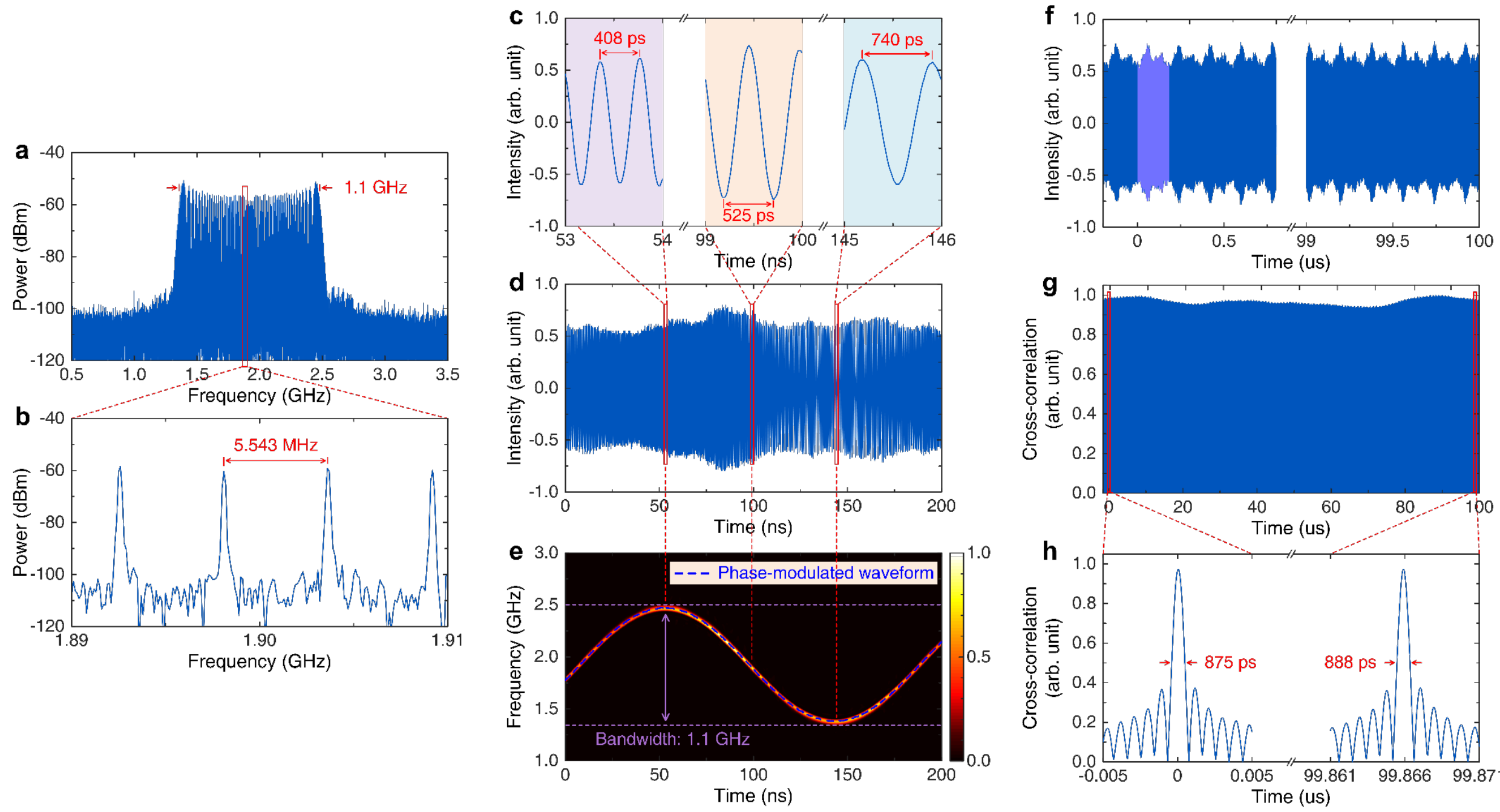

**Fig. 2** Experimental results for the generation of a sinusoidal frequency-modulated microwave signal. **a** Spectrum of the generated sinusoidal frequency-modulated microwave signal with a frequency span of 3 GHz. **b** Spectrum with a frequency span of 20 MHz. **c** The temporal waveforms at the maximum, the middle, and the minimum instantaneous frequencies of the generated chirped microwave signal, respectively. **d** The temporal waveform of the generated chirped microwave signal with a temporal span of 200 ns. **e** The corresponding time-frequency distribution of the temporal waveform in **d**. **f** The generated chirped microwave waveform with a temporal span of about 553 periods used for numerical calculation of the temporal cross-correlation trace. The reference waveform is plotted in light purple. **g** The calculated cross-correlation trace between the reference waveform and the whole waveform plotted in **c**. **h** Two magnifying cross-correlation traces with a null time delay and a delay of about 100 µs, respectively.

The main advantage of the proposed scheme is that only a low-frequency electronic signal $s(t)$ is required, without the need for other high-speed electronics, to manipulate the chirp shape of the generated high-frequency and broadband chirped microwave waveforms. As a reference, the frequency span of $s(t)$ commonly ranges from $f_s$ to a few multiples of $f_s$, resulting in a bandwidth amplification of about $|mf_c / (\pi\Delta f)|$ compared with the bandwidth of the output chirped microwave waveform. Moreover, from another perspective, the proposed configuration provides an equivalent group-velocity dispersion (GVD) to time-stretch an unchirped signal to be a chirped signal. Unlike the traditional GVD induced by optical fiber, the equivalent GVD amount varies in the frequency span of the output bandwidth and can be controlled by $s(t)$. For a typical case of a linearly chirped waveform generation, the equivalent GVD amount (slope of the linear group delay as a function of radial frequency) can be calculated as $\left|\Delta f/\left(2mf_c f_s^2\right)\right|$, which corresponds to propagate through about thousands of kilometers of standard single-mode optical fiber.

**Experiment**. Several experiments based on the fiber-optic platform are implemented to demonstrate the capabilities of the proposed concept for chirped microwave waveform generation, including the tunable central frequency, the controllable output bandwidth, the temporal duration, and the customized chirp shape. The experimental architecture is shown in Fig. 1.

Firstly, we characterize the performance of the proposed scheme through analyzing a sinusoidal frequency-modulated microwave waveform generated from the proposed photonic architecture. In the experiment, the loop FSR of the PML is measured to be 5.563 MHz, corresponding to a round-trip time of 179.8 ns. The phase-modulated signal is a single-tone microwave signal from an AWG (RIGOL), i.e., a sinusoidal signal, and with a repetition rate of 5.543 MHz. As a result, the frequency detuning is equal to 0.02 MHz. Besides, the modulation index is set to its maximum achievable value of $0.72\pi$, which is limited by the high half-wave voltage of the available electro-optic PM (EOSPACE) and the low output power of the AWG. The optical gain provided by an EDFA should be adjusted to slightly larger than 1. In this way, the bandwidth of the chirped optical signal output from the PML is calculated as about 1.1 GHz according to Eq. (8) in the Methods. To ensure the self-oscillation of the chirped optical signals in the PML, the bandwidth of the OBPF is tuned to match the output bandwidth and simultaneously filter out the out-of-band ASE noise. Then, with the help of a home-made optical coherent receiver, a laser (NKT Photonics) with a linewidth less than 100 Hz is used as a local oscillator signal to down-convert the generated chirped optical signal to microwave domain. Finally, a high-speed real-time oscilloscope (Keysight) with a sampling rate of 256 GS/s is employed to capture the temporal waveforms of the generated chirped microwave signal. Figure 2a shows the measured spectrum of the generated sinusoidal frequency-modulated microwave signal. As expected, the output bandwidth is about 1.1 GHz. The detailed spectrum with a span of 20 MHz is displayed in Fig. 2b, where the frequency interval of 5.543 MHz between neighboring teeth, i.e. the repetition rate of the generated chirped microwave signal, is identical to the repetition rate of the phase-modulated signal. Figure 2c exhibits the detailed temporal waveforms at the maximum, the middle, and the minimum instantaneous frequencies of the generated chirped microwave signal. Figure 2d presents the temporal waveform of the generated chirped microwave signal. The temporal waveform is relatively flat and with a temporal period of 180.4 ns. The time-frequency distribution of the generated waveform is obtained by computing the Wigner-Ville distribution of the measured waveform, as shown in Fig. 2e. The frequency span of the time-frequency distribution curve also presents an output bandwidth of 1.1 GHz. As a result, the TBWP of the generated chirped microwave waveform is equal to about 200, limited by the low quality factor of the fiber-based optical loop and the relatively small modulation index. Besides, the phase-modulated waveform, i.e., a single-frequency sinusoidal signal, is also plotted in Fig. 2e to compare it with the chirp shape of the generated chirped signal. For the sake of comparison, the plotted phase-modulated waveform is proportionally scaled to match the amplitude of the time-frequency distribution curve. As consistent with theoretical derivation, the chirp shape of the generated chirped signal is identical to a scaled version of the phase-modulated waveform. In addition, the electro-optic phase modulation is a purely linear process, which guarantees a linearly scaled conversion from the phase-modulated waveform to the chirp shape of the generated chirped microwave waveform. Thus, the proposed scheme provides a low-cost solution for manipulating the time-frequency distribution of the generated chirped microwave signals. Then, to demonstrate the mutual coherence of the generated chirped microwave waveform, a 100 µs-duration temporal waveform of the generated sinusoidal frequency-modulated microwave signal is recorded, corresponding to about 553 periods, as shown in Fig. 2f. A single-period waveform is extracted from the whole waveform (light purple in Fig. 2f) to act as a reference for numerical calculation of the temporal cross-correlation trace. The whole cross-correlation trace presented in Fig. 2g is with a relatively flat profile, and the two cross-correlation peaks at the null time delay and the delay of about 100 µs, as displayed in Fig. 2h, show basically consistent shapes, indicating a high degree of coherence of the generated chirped microwave signal. The pulse width of the cross-correlation peaks is about $\tau = 875$ ps, resulting in a pulse compression ratio, defined as the ratio of the temporal period of the generated chirped signal to $\tau$, of about 206.

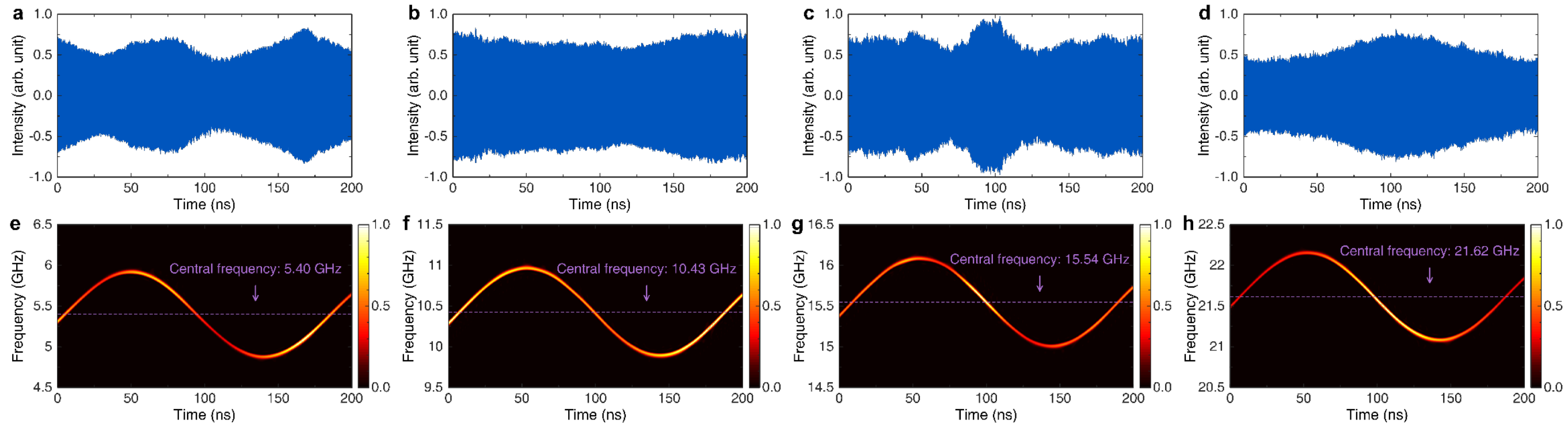

**Fig. 3** Experimental results for tunable central frequency. The central frequencies of the temporal waveforms in **a-d** are equal to 5.40 GHz, 10.43 GHz, 15.54 GHz, and 21.62 GHz, respectively. **e-h** The corresponding time-frequency distributions of the temporal waveforms in **a-d**, calculated by Wigner-Ville distribution.

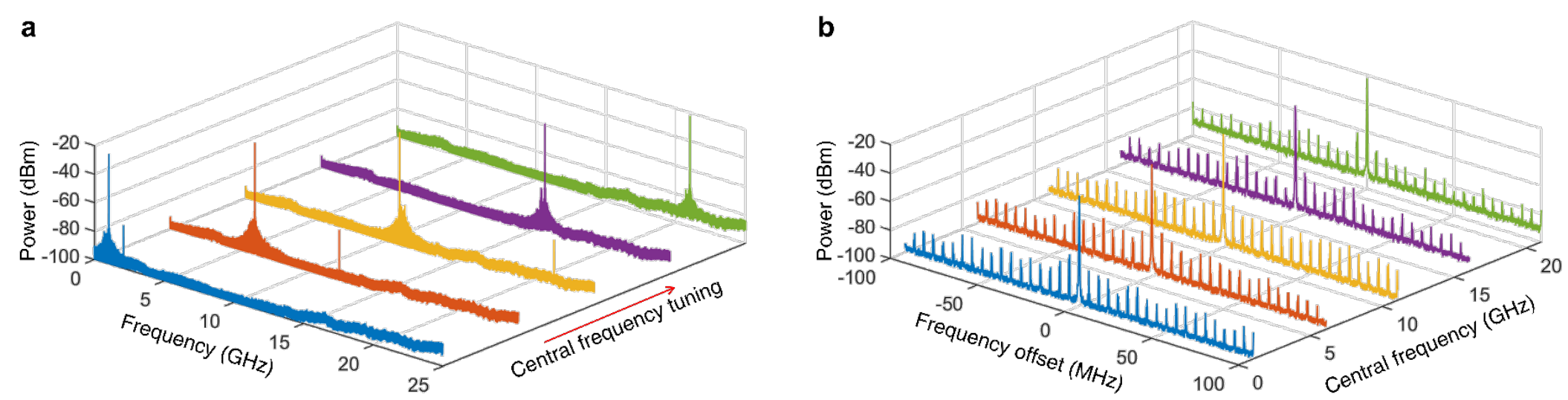

**Fig. 4** Experimental results for tunable central frequency when the PML operates at free-running state. **a** The frequency tuning of the output single-tone microwave signals by changing the frequency of the local oscillator laser. **b** Zoom-in views of **a** at the central frequencies.

Then, we demonstrate the ability of the proposed scheme to tune the central frequency of the generated chirped microwave signals through changing the frequency of the laser used as a local oscillator signal. Figure 3 shows four generated sinusoidal frequency-modulated microwave signals with an identical bandwidth but with different central frequencies of 5.40 GHz, 10.43 GHz, and 21.62 GHz, respectively. The variation of the output central frequency is equal to that of the single-frequency laser. The distinction between the temporal envelopes of the four chirped microwave signals is caused by the frequency response difference of the electronic detection devices in different frequency bands. To further investigate the central frequency tuning process, the PML operates at a free-running state, namely, without the phase modulation operation, to observe the frequency tuning of the single-tone microwave signal. Figure 4a displays the spectrums of five single-tone microwave signals generated by down-converting the optical signal to microwave domain with the help of five different local oscillator frequencies. The $2^{nd}$-order harmonics shown in Fig. 4a are produced by the nonlinear photoelectric conversion due to the large input optical power. The central frequency can be tuned in a frequency span of about 21 GHz, limited by the bandwidth of the optical coherent receiver. The zoom-in views of Fig. 4a at the central frequencies are shown in Fig. 4b. Owing to the mode competition, the output of the free-running PML only contains a single-tone signal and is with a side-mode suppression ratio larger than 40 dB, which gives evidence for supporting the mode competition deduction in the Methods. Alternatively, the central frequency of the generated chirped microwave signals can also be tuned by adjusting that of the OBPF.

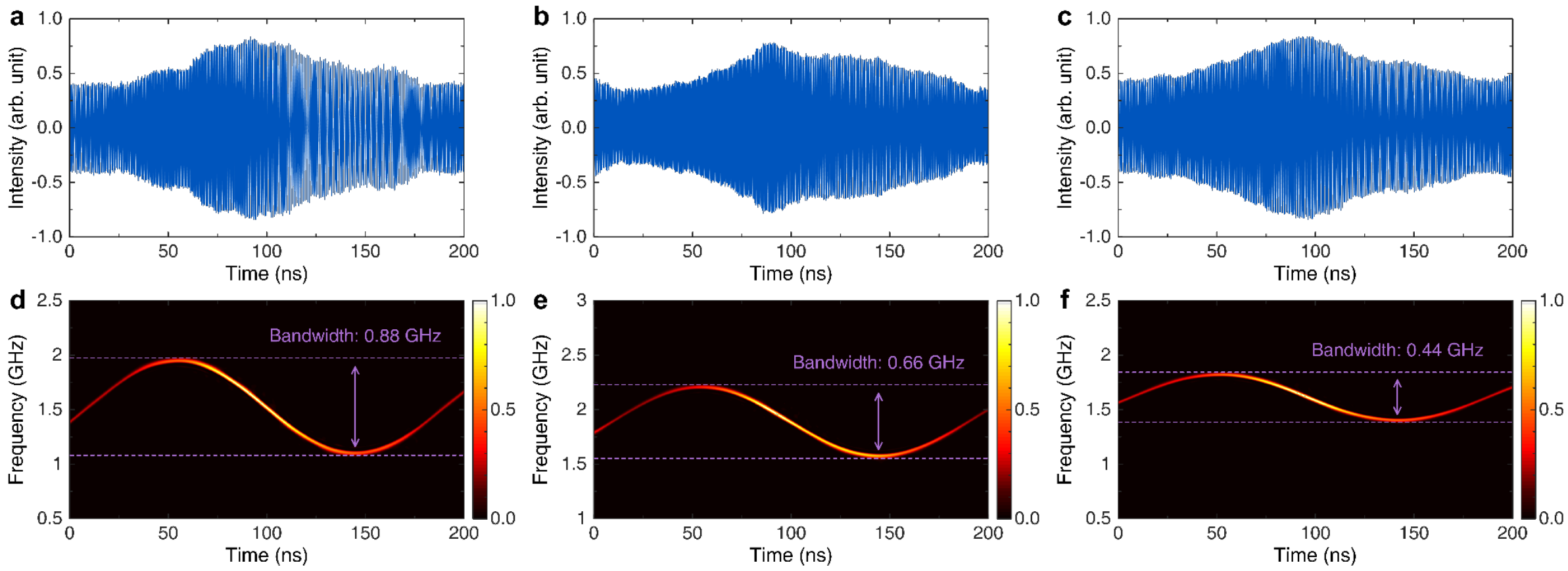

**Fig. 5** Experimental results for controllable bandwidth through changing the modulation index. The modulation indexes applied for generating the temporal waveforms in **a-c** are equal to 0.570$\pi$, 0.428$\pi$, and 0.285$\pi$, respectively. **d-f** The corresponding time-frequency distributions of the temporal waveforms in **a-c**, calculated by Wigner-Ville distribution.

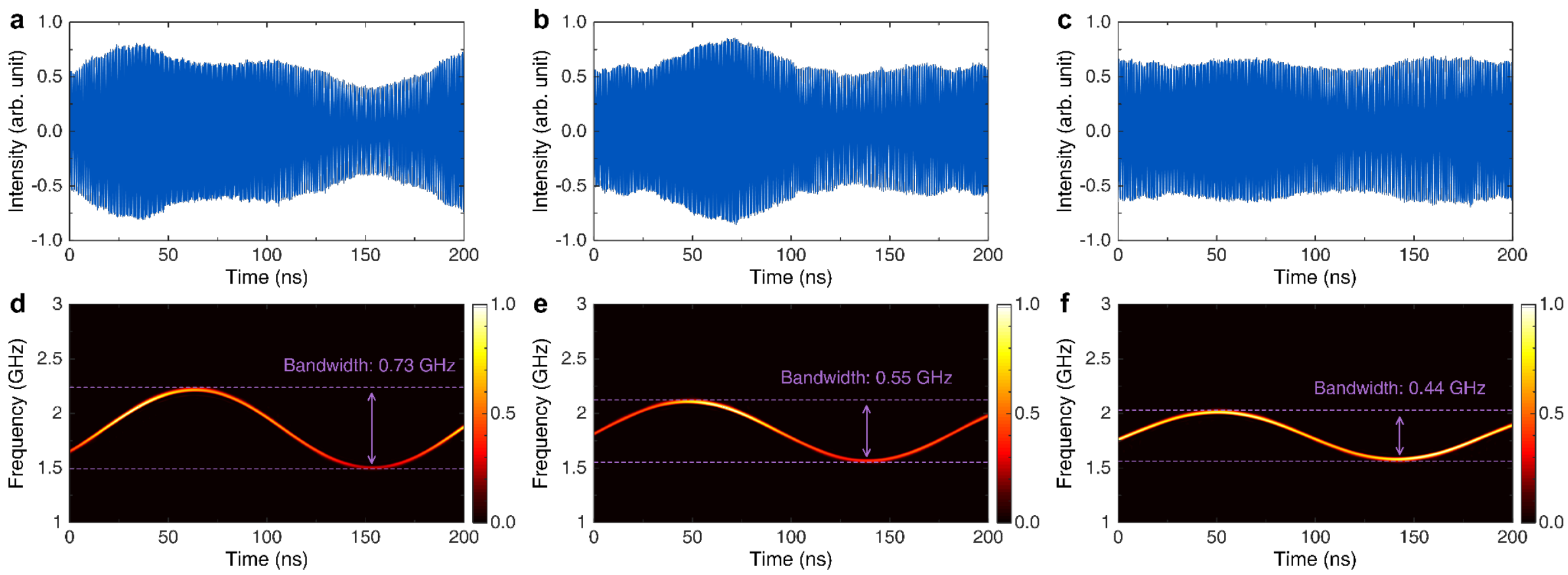

**Fig. 6** Experimental results for controllable bandwidth through changing the frequency detuning. The frequency detuning employed for generating the temporal waveforms in **a-c** are equal to 0.03 MHz, 0.04 MHz, and 0.05 MHz, respectively. **d-f** The corresponding time-frequency distributions of the temporal waveforms in **a-c**, calculated by Wigner-Ville distribution.

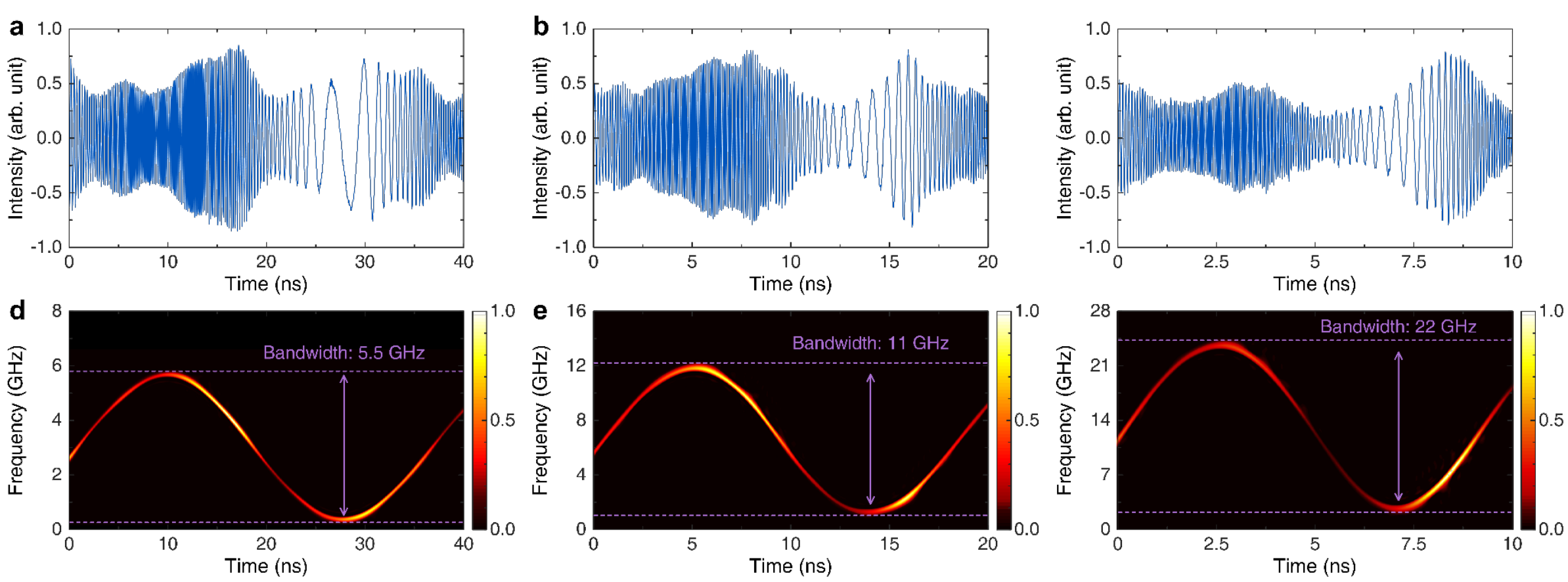

**Fig. 7** Experimental results for controllable bandwidth through changing the repetition rate of the phase-modulated waveform. The repetition rates of the phase-modulated waveforms used for generating the temporal waveforms in **a-c** are equal to 27.793 MHz, 55.610 MHz, and 111.240 MHz, respectively. **d-f** The corresponding time-frequency distributions of the temporal waveforms in **a-c**, calculated by Wigner-Ville distribution.

The bandwidth of the generated chirped signals is another controllable parameter in the proposed architecture. According to Eq. (8), without changing the loop length of the PML, i.e., a fixed $f_c$, the output bandwidth depends on three parameters, including the modulation index, the frequency detuning, and the repetition rate of the phase-modulated waveform. Figure 5 shows three sinusoidal frequency-modulated microwave signals with different bandwidth and their corresponding time-frequency distributions, where the modulation indexes are equal to $0.570\pi$, $0.428\pi$, and $0.285\pi$, respectively. As expected, the ratio between the bandwidths of the generated chirped microwave signals is equal to 4:3:2, i.e., 0.88 GHz : 0.66 GHz : 0.44 GHz. The similar envelopes of the temporal waveforms shown in Fig. 5a-c indicate that the increase of the modulation index makes no distortion on the output temporal waveform. In another way, the output bandwidth can also be controlled by adjusting the frequency detuning, as displayed in Fig. 6. The output bandwidth is inversely proportional to the frequency detuning, which is consistent with Eq. (8) in the Methods. Without the variation of the frequency detuning, the repetition rate of the phase-modulated waveform can be increased by times to enhance the output bandwidth. Figure 7 shows the generated temporal waveforms and their corresponding time-frequency distributions when the repetition rates of the phase-modulated signals are increased to five times, ten times, and twenty times of that applied in Fig. 2. In this case, the resulting output bandwidths will be proportionally increased to 5.5 GHz, 11 GHz and 22 GHz. Here, the maximum output bandwidth is limited by the 3-dB bandwidth of the optical coherent receiver. Besides, as shown in Fig. 7, the temporal durations of the chirped microwave signals are proportionally decreased to 36 ns, 18 ns and 9 ns due to the diminution in modulation periods.

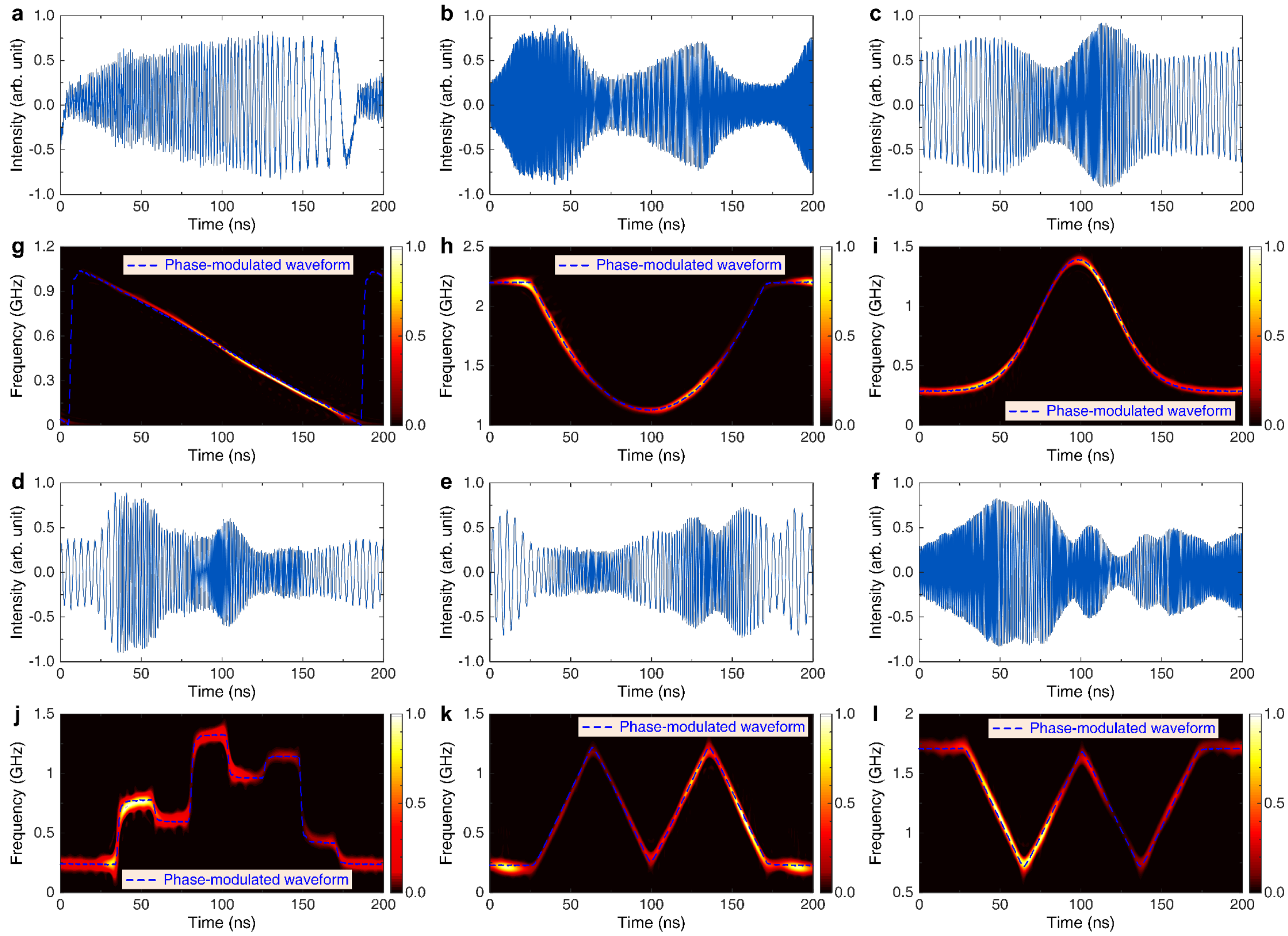

**Fig. 8** Experimental results for customized chirp shapes. The generated chirped signals with temporal waveforms shown in **a-f** are linearly frequency-modulated signal, quadratically frequency-modulated signal, gaussian-pulse-shaped frequency-modulated signal, Costas-coded frequency-hopping signal with a 6-bit Costas code sequence of [3 2 6 4 5 1], 'M'-shaped frequency-modulated signal, and 'W'-shaped frequency-modulated signal, respectively. **g-l** The corresponding time-frequency distributions of the temporal waveforms in **a-f**, calculated by Wigner-Ville distribution.

The most remarkable advantage provided by the proposed scheme is that the chirp shape of the generated chirped microwave waveform, i.e., time-frequency distribution, is identical to a scaled version of the phase-modulated waveform. In other words, the time-frequency distribution of the generated chirped microwave waveforms can be arbitrarily customized through simply changing the phase-modulated waveform, without other complex operations. Besides, the bandwidth ratio of the generated chirped microwave signal to the phase-modulated signal is generally larger than several hundred, which means that the time-frequency distribution of a high-frequency and broadband chirped microwave signal can be manipulated by a low-frequency waveform. Figure 8 exhibits the temporal waveforms and the time-frequency distributions of eight generated chirped microwave signals, including fundamental chirped signals as well as the customized chirped signals with time-frequency distributions identical to the letters 'M' and 'W'. Thereinto, all the chirp shapes of the output signals are identical to a scaled form of the phase-modulated waveforms, which shows a robust manipulation ability in the time-frequency distribution.

## Discussion

We have proposed, theoretically analyzed and experimentally demonstrated a novel concept for generating broadband chirped microwave waveforms with arbitrarily chirp shapes by using an extremely simple architecture. Our proposed concept is based on the recirculating phase modulation in an oscillating loop/cavity, where the modulation period is slightly deviated from the loop/cavity round-trip time. The proof-of-concept experiments are carried out in a fiber-optic platform containing a simple recirculating phase-modulated optical fiber loop and driven by low-speed electronics, avoiding the use of pulse shapers, frequency-scanning filters, large dispersion mediums, or high-frequency electronics, which generally appear in the vast majority of the approaches reported so far. As a result, the proposed concept for the chirped microwave signal generation offers a remarkable improvement in terms of the system cost, the structural complexity, and the corresponding SWaP, which makes it more feasible for the application in real world. Besides, the proposed concept can also be applied in other platforms such as optoelectronic oscillators (OEOs) and optical microresonators, where a more distinguished performance can be exhibited since the higher quality factor is provided by them.

The chirped microwave waveforms generated from the proposed photonic architecture can be flexibly reconfigurable in the central frequency, the bandwidth, the time duration, and the time-frequency distribution. The tuning range of the central frequency and the bandwidth reported here are exclusively limited by the available bandwidth of the optical coherent receiver, which can be easily enhanced to exceed a few hundred GHz by employing a faster PD such as a UTC-PD. The time duration of the generated chirped signals is identical to the phase-modulated period, and its maximum is limited by the loop round-trip time. Hence, the chirped microwave signals with a time duration in the μs range are easily generated in a longer phase-modulated optical loop. Accordingly, the generation of shorter chirped microwave waveforms is through simply increasing the repetition rate of the phase-modulated signals, without the variation of the loop length. In addition, the most extraordinary feature offered by the proposed concept is that the chirp shape of the generated chirped microwave signal is identical to a scaled form of the phase-modulated waveform, which gives us an extremely simple method to manipulate the chirp shape by adjusting the phase-modulated waveform. The cross-correlation trace of the output chirped microwave signals indicates a highly mutual coherence in a time span exceeding 100 μs. This characteristic is greatly prominent for many practical applications such as the pulse compression radar system. The coherent time larger than 100 μs enables the detection range of the radar systems to exceed tens of kilometers. The TBWP of about 200 in the experiment is restricted by the limited modulation index and quality factor of the fiber-based optical loop. With the help of the cascaded phase modulation structure and the photonic integration technology, the TBWP of the generated chirped microwave signals can be enhanced by several orders of magnitude. The demonstrated performance and the predictable potential development of the proposed concept show an excellent advancement in the generation of chirped microwave waveforms that are suitable for numerous applications such as ranging, wireless communication and biomedical imaging.

## Data availability

The data that support the findings of this study are available from the corresponding author upon reasonable request.

## Acknowledgements
This work was supported by the National Natural Science Foundation of China (61927821), and the Fundamental Research Funds for the Central Universities (ZYGX2020ZB012).

## Author contribution
W.Q.L. and Z.Y.Z. conceived the project. W.Q.L. carried out the experiments. H.T. and Z.W.F. performed the numerical simulations. L.J.Z., Z.Z., and Y.W. Z. conducted the theoretical analysis. W.Q.L., Z.Y.Z., and Y.L. carried out the data analysis. W.Q.L. wrote the manuscript with contributions from all authors. Y.L. finalized the paper. Z.Y.Z and Y.L. supervised the project.

## Competing interests
The authors declare no competing interests.